\documentclass[12pt,final]{amsart}

\usepackage[margin=1in]{geometry}

\usepackage{amssymb}

\newtheorem{lemma}{Lemma}[section]

\newtheorem{theorem}{Theorem}[section]

\theoremstyle{definition}
\newtheorem{definition}{Definition}[section]
\newtheorem{corollary}{Corollary}[section]
\newtheorem{remark}{Remark}[section]

\newcommand{\Z}{\ensuremath{{\mathbb Z}}}
\newcommand{\N}{\ensuremath{{\mathbb N}}}

\newcommand{\R}{\ensuremath{\mathbb R}}
\newcommand{\C}{\ensuremath{\mathbb C}}

\newcommand{\Prj}{\ensuremath{\mathbb P}}

\newcommand{\Hk}{\ensuremath{\mathcal{H}}_k}
\newcommand{\U}{\ensuremath{\mathcal{U}}}

\newcommand{\tr}{\ensuremath{{\mathrm {Tr}}}}
\newcommand{\ndom}{\ensuremath{{\mathrm {End}}}}

\author{Tatyana Barron}
\address{T. Barron, Department of Mathematics, 
University of Western Ontario, 
London, Ontario N6A 5B7, Canada }
\email{tatyana.barron@uwo.ca}

\author{Timothy Pollock}
\address{T. Pollock, Department of Mathematics,
University of Western Ontario, 
London, Ontario N6A 5B7, Canada }
\email{tpolloc4@uwo.ca}

\thanks{Research is supported in part 
by the Natural Sciences and Engineering Research Council of Canada}

\title{K\"ahler quantization and entanglement}

\begin{document}

\maketitle

\noindent {\bf Abstract.}  
For a very ample line bundle $L$ on a compact connected complex manifold $X$, with a real structure,
we discuss entanglement properties of certain   
sequences of vectors in tensor products of spaces of holomorphic sections of powers of $L$.

\


\section{Introduction}

Let $L\to X$ be a very ample holomorphic hermitian line bundle on a compact connected complex manifold $X$. 
Paper \cite{borthwick:95} discusses a certain sequence $(u_k)$  of holomorphic sections of $L^{\otimes k}$   
associated to a Legendrian submanifold of the unit circle bundle $Z$ in $L^*$  
(projection of this submanifold to $X$ is a Lagrangian submanifold of $X$ satisfying a Bohr-Sommerfeld 
condition). The K\"ahler form on $X$ and the contact form on $Z$ are obtained from the hermitian metric on $L$.   
It is customary to interpret $k$ as $\frac{1}{\hbar}$.

Associating semiclassical states to Lagrangian submanifolds of a symplectic manifold is an idea that arises 
in many contexts. 
A partial list of references where various versions of this appear    
 includes \cite{bates:97, borthwick:95, burns:10, deber:06, gorod:01, jeffrey:92, 
paoletti:08} 
and three papers (co-)authored by the first author of this paper: \cite{alluhaibi:16, foth:02, foth:08}.  

Similar ideas are applicable to isotropic submanifolds \cite{guillemin:15}.

When the states are associated to submanifolds of a product manifold, we address 
the question of entanglement, that comes from quantum information theory. We work with pure states. 
Entanglement measures,  
such as the entropy of entanglement, are defined for vectors in a tensor product of two or more 
Hilbert spaces. In our setting the Hilbert space comes from K\"ahler quantization. We find that when the Lagrangian 
submanifold $\Lambda$ of $X\times X$ is $X$, embedded antidiagonally (for this $X$ needs to have a real structure), 
the sequence of normalized vectors in spaces of holomorphic sections of the line bundles, which is associated 
to $\Lambda$, is a sequence of maximally entangled vectors (Theorem \ref{entropyth}). 
We start with a very ample line bundle $L\to X$ on a compact connected complex manifold $X$, with a real structure. 
Presence of the real structure makes $\Lambda\cong X$ a Bohr-Sommerfeld submanifold of $X\times X$. We construct a sequence 
$\{ u_{\Lambda}^{(k)}\}$, $k=1,2,3,...$ (a sequence of holomorphic sections of
powers of the quantum line bundle on $X\times X$) by  
applying a standard procedure of associating states to a Bohr-Sommerfeld Lagrangian submanifold. 
Since the space of holomorphic sections of the $k$-th power of the quantum line bundle on $X\times X$ is isomorphic to 
$H^0(X,L^{\otimes k})\otimes H^0(X,L^{\otimes k})$ (Remark \ref{remisom}), it is natural to inquire about the entanglement properties 
of $u_{\Lambda}^{(k)}$.  
We show that the entropy of entanglement 
$\nu (\frac{u_{\Lambda}^{(k)}}{||u_{\Lambda}^{(k)}||})=\ln {\mbox{dim}} H^0(X,L^{\otimes k})$, 
which means that $\{ \frac{1}{||u_{\Lambda}^{(k)}||}u_{\Lambda}^{(k)} \}$ is a sequence of maximally entangled vectors. 
For this sequence we also describe 
how the entropy of entanglement is related to the minimum distance to a separable vector (Corollary \ref{cormindist}).  
As an example, when $X=\Prj ^1$, with the Fubini-Study form, 
the entropy of entanglement of the $k$-th vector is $\ln(k+1)$.This is in Section \ref{sphere}. In Section \ref{torus} we consider 
the case when $X$ is a $2$-dimensional torus. 

We point out that a similar line of thought can be applied to a submanifold $\Lambda'$ of $\Lambda$. 
As an example, we calculated the entropy of entanglement 
for the sequence of sections associated to a circle in $\Prj ^1$. These vectors are entangled but not maximally 
entangled, for each fixed $k$ or as $k\to\infty$. This calculation is in Section \ref{sphere}.

Real structures appear often in mathematical physics. See \cite{doran:15} as one among many papers on string theory 
where real structures play a substantial role, also see \cite[Chapter 2]{manin:88} for a discussion in the context of field theory. 
In K\"ahler quantization, which is relevant to this paper, existence of a real structure 
on the K\"ahler manifold (in the sense of the definition in Section \ref{sectantidiag}) can be translated, 
intuitively, into having quantum states that are their own antiparticles, analogously to Majorana fermions.

There is extensive literature on representation-theoretic aspects of entanglement.  
We note papers 
\cite{huckleberry:13} and \cite{sawicki:11} where the point of view taken is that on the tensor product 
of two Hilbert spaces (or of finitely many Hilbert spaces)  
entanglement 
should be studied 
on the orbits of the action of the unitary group.

In this paper we consider the questions about separability and entanglement combined with concepts 
from K\"ahler quantization.

\section{Entropy of entanglement for sequences of sections of line bundles}

\subsection{Preliminaries}
Let $(X,\omega )$ be a compact connected $n$-dimensional K\"ahler manifold ($n\ge 1$). 
Assume that $\frac{\omega}{2\pi}$ is an integral K\"ahler form.  
There is a  holomorphic hermitian line bundle $L\to X$ such that the curvature of the Chern connection 
is equal to $-i \omega$. For a positive integer $k$  
denote $\Hk =H^0(X,L^{\otimes k})$ (the space of holomorphic sections of $L^{\otimes k}$).  
It is a finite-dimensional Hilbert space and  $d_k={\mbox{dim}} \ \Hk \sim {\mathrm{vol}} (X) k^n+O(k^{n-1})$, 
as $k\to\infty$, with ${\mathrm{vol}} (X)=\int\limits_X\frac{c_1(L)^n}{n!}$ 
(this is a standard fact that can be found, for example, in \cite[Chapter 4]{ma:07}).   
The inner product $h_{L_x}(.,.)$ on $L_x$, $x\in X$, induces the inner product $h_{L^*_x}(.,.)$ on $L^*_x$,  
the inner product  $h_{L_x\otimes L_x}(.,.)$ on $L_x\otimes L_x$ and so on. 
Every element $\xi$ of $L^*_x$ can be represented as $h_{L_x}(.,v)$ for some $v\in L_x$, with $v$ and $\xi$ of the same norm.  
If $\eta\in L^*_x$ is $h_{L_x}(.,w)$ for $w\in L_x$, then $h_{L_x^*}(\xi,\eta)=h_{L_x}(w,v)$. 
For $v,w,v',w'\in L_x$ $h_{L_x\otimes L_x}(v\otimes v',w\otimes w')=h_{L_x}(v,w)h_{L_x}(v',w')$. 
The Hermitian metric 
on $L$ provides an inner product on $\Hk$. Let $dV=\frac{\omega^n}{n!}$. 
For $s_1,s_2\in \Hk$ 
$$
\langle s_1,s_2\rangle =\int\limits_X h_{L_x^{\otimes k}}(s_1(x),s_2(x))dV(x).
$$ 
$\Hk\otimes \Hk$ is a tensor product of Hilbert spaces, 
with the inner product $\langle u\otimes u', w\otimes w'\rangle = \langle u, w\rangle\langle u',w'\rangle$ 
for $u,u',w,w'\in \Hk$ and extended by linearity. 
\begin{definition}
A vector $v\in \Hk \otimes \Hk$ is called {\it separable} or {\it decomposable} if it is of the form 
$v_1\otimes v_2$, $v_1, v_2\in \Hk$. 
\end{definition}
Let $\varphi_1^{(k)}$,...,$\varphi_{d_k}^{(k)}$ be an orthonormal basis in $\Hk$. Then 
$\varphi_j^{(k)}\otimes \varphi_l^{(k)}$, $j,l=1,...,d_k$, is an orthonormal basis in 
$\Hk\otimes\Hk$.  
Let $\pi_j$, $j=1,2$, denote the projections $X\times X\to X$ onto the first and second factor respectively.  
The line bundle $L$ is ample and for a sufficiently large $k_0\in \N$ the line bundle 
$L^{\otimes k}$ is 
very ample for $k\ge k_0$. 
We shall assume that $\omega$ is such that $L$ is very ample (this can be achieved by replacing 
$\omega $ by   $\omega'=k_0\omega$). 
\begin{remark}
\label{remisom}
$H^0(X\times X,\pi_1^*L^{\otimes k}\otimes \pi_2^*L^{\otimes k})\cong \Hk\otimes \Hk$ for all $k$ (see a proof in \cite{barron:17}). 
\end{remark}
\begin{definition}
The {\it entropy of entanglement} for a vector $v$ 
in $\Hk\otimes \Hk$, of norm $1$, 
is 
$$
\nu(v)=-\sum\limits_{j=1}^{d_k}\lambda_j\ln \lambda_j,
$$ 
where  $\lambda_1$,...,$\lambda_{d_k}$ 
are the eigenvalues of $\tr_2(v\bar{v}^T)$, and $\tr _2: \ndom (\Hk \otimes \Hk) \to \ndom (\Hk )$ is defined 
by $\tr _2(A\otimes B ) =\tr (B )A$ for $A,B \in \ndom (\Hk)$ and extended by linearity. 
The convention $0\ln 0 =0$ is used. 
\end{definition}
In the definition above 
the minimum value of $\nu(v)$ is zero, and   $\nu(v)=0$ 
if and only if $v$ is separable.  
The maximum value of $\nu(v)$ is $\ln d_k$. It is not hard to see that $\nu(v)=\ln d_k$ if and only if 
$\lambda_1=...=\lambda_{d_k}=\frac{1}{d_k}$. 
\begin{definition}
A vector $v\in \Hk\otimes \Hk$, of norm $1$, 
is called {\it maximally entangled} if $\nu(v)=\ln d_k$. 
\end{definition}
The value of $\nu(v)$ characterizes how entangled  (or how "not separable") 
$v$ is. 
\begin{definition}
The {\it Schmidt decomposition} for $v\in \Hk\otimes \Hk$ is a representation of $v$ as 
$v=\sum\limits_{j=1}^{d_k}\alpha_j^{(k)}\psi_j^{(k)}\otimes\chi_j^{(k)}$, where $\{ \psi_j^{(k)} \}$ is an orthonormal basis 
of $\Hk$, $\{ \chi_j^{(k)} \}$ is another orthonormal basis 
in  $\Hk$, and $\alpha_1^{(k)}\ge ...\ge \alpha_{d_k}^{(k)}\ge 0$ are real numbers. 
\end{definition}
Such a representation 
always exists and 
$$
\tr _2(v\bar{v}^T)=\begin{pmatrix} (\alpha_1^{(k)})^2 & & & \\ 
& (\alpha_2^{(k)})^2 & & \\
 & &  ... & \\
 & & & (\alpha_{d_k}^{(k)})^2\end{pmatrix}.
$$
All this is standard material,  - a reference is, for example, \cite[Sections 9.2, 15.3]{bengt:06}. 
Note that if $v$ is of norm $1$ then, of course, $\sum\limits_{j=1}^{d_k}(\alpha_{j}^{(k)})^2=1$.

For $u,v\in \Hk\otimes \Hk$ denote by $D(v,u)$ the distance $D(v,u)=\sqrt{ \langle v-u,v-u\rangle}$. 
\begin{lemma}
\label{closestsepv}
For $v\in \Hk\otimes \Hk$ the minimum value of $D(v,u)$, for $u$ separable, is $\sqrt{\sum\limits_{j=2}^{d_k}(\alpha_j^{(k)})^2}=D(v,u_s)$, where 
$u_s=\alpha_1^{(k)} \psi_1^{(k)}\otimes\chi_1^{(k)}$.  
\end{lemma}
\begin{remark}
Lemma \ref{closestsepv} seems to be known (see e.g. \cite{lockhart:02}). However, we have not seen the statement 
in this exact form, and we provide a proof. 
\end{remark}
\noindent {\bf Proof.} 
Let $v=\sum\limits_{j=1}^{d_k}\alpha_j^{(k)}\psi_j^{(k)}\otimes\chi_j^{(k)}$ be the Schmidt decomposition for $v$ 
and let $u_1=\sum\limits_{j=1}^{d_k}a_j\psi_j^{(k)}$, $u_2=\sum\limits_{l=1}^{d_k}b_l\chi_l^{(k)}$ be two vectors in $\Hk$. 
$$
D(v,u_1\otimes u_2)^2=\langle v-u_1\otimes u_2,v-u_1\otimes u_2\rangle =
\sum\limits_{j=1}^{d_k}(\alpha_j^{(k)})^2+\sum\limits_{j=1}^{d_k}\sum\limits_{l=1}^{d_k}|a_jb_l|^2-
\sum\limits_{j=1}^{d_k}\alpha_j^{(k)}(a_jb_j+\bar{a}_j\bar{b}_j), 
$$
$$
D(v,u_s)^2=\sum\limits_{j=2}^{d_k}(\alpha_j^{(k)})^2, 
$$
and it remains to show:
$$
D(v,u_1\otimes u_2)^2\ge D(v,u_s)^2 
$$
or, equivalently, 
$$
(\alpha_1^{(k)})^2+\sum\limits_{j=1}^{d_k}\sum\limits_{l=1}^{d_k}|a_jb_l|^2-
\sum\limits_{j=1}^{d_k}\alpha_j^{(k)}(a_jb_j+\bar{a}_j\bar{b}_j)\ge 0.
$$
Using that $\alpha_1^{(k)}\ge \alpha_j^{(k)}\ge 0$ and the Cauchy-Schwarz inequality, we get: 
$$
(\alpha_1^{(k)})^2+\sum\limits_{j=1}^{d_k}|a_j|^2\sum\limits_{l=1}^{d_k}|b_l|^2-
\sum\limits_{j=1}^{d_k}\alpha_j^{(k)}(a_jb_j+\bar{a}_j\bar{b}_j) 
$$
$$
\ge (\alpha_1^{(k)})^2+(a_1b_1+...+a_{d_k}b_{d_k})(\bar{a}_1\bar{b}_1+...+\bar{a}_{d_k}\bar{b}_{d_k})-
\sum\limits_{j=1}^{d_k}\alpha_1^{(k)}(a_jb_j+\bar{a}_j\bar{b}_j)
$$
$$
=|\alpha_1^{(k)}-(a_1b_1+...+a_{d_k}b_{d_k})|^2\ge 0.
$$
$\Box$
\begin{corollary}
\label{cormaxent}
If $v$ is maximally entangled then
$$
D(v,u_s^{(k)})=\sqrt{1-e^{-\nu(v)}}
$$
\end{corollary}
\noindent {\bf Proof.} 
If $v$ is maximally entangled then 
$D(v,u_s^{(k)})=\sqrt{\frac{d_k-1}{d_k}}$, and the statement follows.
$\Box$

\begin{definition}
The {\it coherent vector} $u_\xi^{(k)}\in \Hk$ (for $x\in X$, $0\ne \xi\in L_x^*$) 
is the unique element of $\Hk$ with the property 
\begin{equation}
\label{defcohpoint}
\langle s,u_\xi^{(k)}\rangle=\xi^{\otimes k}(s(x))
\end{equation} 
for any $s\in\Hk$. 
The {\it coherent state} associated to $u_\xi^{(k)}$ is $\pi (u_\xi^{(k)})$, where 
$\pi :\Hk -\{ 0\}\to \Prj(\Hk)$ is the projectivization map. 
\end{definition}
\begin{remark}
As it is clear from the equality (\ref{defcohpoint}), 
multiplying $\xi$ by a complex number $\alpha\ne 0$ leads to changing   $u_\xi^{(k)}$ by a factor of $\bar{\alpha}^k$.
\end{remark}
This kind of coherent states originates, in somewhat different ways, from work of  
Berezin and of Rawnsley   \cite{schlich:10}, \cite{rawnsley:77}.

We observe that $u_\xi^{(k)}=\sum\limits_{j=1}^{d_k}\overline{\xi^{\otimes k}(\varphi_j^{(k)}(x))}\varphi_j^{(k)}$ 
(because, if we represent this vector as $u_\xi^{(k)}=\sum\limits_{j=1}^{d_k} c_j\varphi_j^{(k)}$, where $c_j$ are  coefficients, then  
for each $l\in \{1,...,d_k\}$ 
$\langle \varphi_l^{(k)},u_{\xi}^{(k)}\rangle = \bar{c}_l= 
\xi^{\otimes k}(\varphi_l^{(k)}(x))$). 

For $x,y\in X$, $0\ne \xi\in L_x^*$, $0\ne \eta\in L_y^*$ we will denote by  
$u_{\xi,\eta}^{(k)}\in \Hk\otimes \Hk$ the element of $\Hk\otimes \Hk$ 
with the property 
$$
\langle w_1\otimes w_2 ,u_{\xi,\eta}^{(k)}\rangle=
\xi^{\otimes k}(w_1(x))\eta^{\otimes k}(w_2 (y))
$$
for all $w_1,w_2\in\Hk$.  
It is not hard to see that 
$$
u_{\xi,\eta}^{(k)}=u_\xi^{(k)}\otimes u_\eta^{(k)}=\sum\limits_{j=1}^{d_k}\sum\limits_{l=1}^{d_k}
\overline{\xi^{\otimes k}(\varphi_j^{(k)}(x))\eta^{\otimes k}(\varphi_l^{(k)}(y))}
\varphi_j^{(k)}\otimes \varphi_l^{(k)}.
$$  
We will consider $X\times X$ with the symplectic form $\Omega=\pi_1^*\omega +\pi_2^*\omega$. 
Denote the Chern connection in ${\mathcal{L}}=\pi_1^*L\otimes \pi_2^*L \to X\times X$ by $\nabla$ 
(note that its curvature is $-i\Omega$). 

Suppose  $\Lambda$ is a Lagrangian submanifold of $X\times X$ satisfying the Bohr-Sommerfeld condition 
(i.e. $\nabla$ over $\Lambda$ is trivial \cite{burns:10, eliashberg:95}).  
\begin{remark}
Equivalently, the Bohr-Sommerfeld condition may be stated as a requirement that ${\mathcal{L}}\Bigr | _{\Lambda}$ 
has a nonzero covariant constant section \cite{burns:10, jeffrey:92}.
\end{remark}
Let $\nabla^*$ be the hermitian connection in ${\mathcal{L}} ^*$ induced by $\nabla$. 
Let $\tau$ be a nonvanishing covariant constant section of ${\mathcal{L}}^*\Bigr | _{\Lambda}$. 
For $(x,y)\in \Lambda$ we will denote by 
$u_{\tau (x,y)}^{(k)}$ the  element of $\Hk\otimes \Hk$ 
with the property 
$$
\langle w_1\otimes w_2,u_{\tau (x,y)}^{(k)}\rangle=
\tau(x,y)^{\otimes k}(w_1(x)\otimes w_2(y))
$$
for all $w_1,w_2\in\Hk$. 
We observe that  
$$
u_{\tau (x,y)}^{(k)}=\sum\limits_{j=1}^{d_k}\sum\limits_{l=1}^{d_k}\overline{\tau(x,y)^{\otimes k}(\varphi_j^{(k)}(x)\otimes \varphi_l^{(k)}(y))}\varphi_j^{(k)}\otimes \varphi_l^{(k)}.
$$ 
Let $d\mu$ be an $n$-form on $\Lambda$, with $\int\limits_{\Lambda} d\mu >0$.
We will denote by $u_{\Lambda}^{(k)}$ the element of $\Hk \otimes \Hk$ with the property
$$
\langle w_1\otimes w_2,u_{\Lambda}^{(k)}\rangle = \int\limits_{\Lambda}
\tau(x,y)^{\otimes k}( w_1(x)\otimes w_2(y)) \ d\mu (x,y)
$$ 
for all $w_1,w_2\in \Hk$.  Clearly 
$$
u_{\Lambda}^{(k)}=\sum\limits_{j=1}^{d_k}\sum\limits_{l=1}^{d_k}
\varphi_j^{(k)}\otimes \varphi_l^{(k)} \int\limits_{\Lambda}
\overline{\tau(x,y)^{\otimes k} (\varphi_j^{(k)}(x)\otimes \varphi_l^{(k)}(y))}d\mu(x,y).
$$ 

\subsection{$X$ antidiagonally embedded into $X\times X$}
\label{sectantidiag}
Now suppose $X$ has a real structure (i.e.  
an antiholomorphic involution $\sigma:X\to X$). We require that, moreover, $\sigma^*\omega =-\omega$
and 
\begin{equation}
\label{sigmareq}
h_{L_{\sigma(x)}^{\otimes k}}(s_1(\sigma(x)),s_2(\sigma(x)))=\overline{h_{L_x^{\otimes k}}(s_1(x),s_2(x))}
\end{equation}
for all $x\in X$, $k\in \N$, $s_1,s_2 \in\Hk$. This implies that  
$\sigma$ extends to  
a norm-preserving bundle map $L^*\to L^*$ that is antilinear on the fibers, 
for any $x\in X$, $\sigma$  identifies $L_{\sigma(x)}^*$ with $\overline{L_x^*}\cong L_x$, and so 
$\sigma^*L\cong L^*$.

Define $\Lambda=\{ (x,y)\in X\times X| \ y=\sigma(x)\}$ and 
$\iota:X\overset{\cong}{\to} \Lambda$, $x\mapsto (x,\sigma(x))$. 
We have: $\iota^*\pi_1^*L\cong L$, $\iota^*\pi_2^*L\cong L^*$, 
$\nabla$ restricted to  ${\mathcal{L}}\Bigr |_{\Lambda}$ 
is trivial. To show that $\nabla$ is trivial over $\Lambda$, 
choose a local holomorphic frame $e(x)$ for $L$ and observe that 
over $\Lambda \cong X$ the connection form for $\nabla$ is 
$$
\iota^* \partial \log \Bigl ( h_{L_x}(e(x),e(x))h_{L_{\sigma(x)}}(e(\sigma(x)), e(\sigma(x)) ) \Bigr ) \Bigr | _{\Lambda}
$$
$$
=
  \partial \log h_{L_x}(e(x),e(x))+\bar{\partial }\log  h_{L_x}(e(x),e(x))=d \log  h_{L_x}(e(x),e(x)).
$$
Since the connection form is exact, the connection is trivial.

Let $\tau$ be a (nonvanishing, covariant constant) section of  ${\mathcal{L}}^*\Bigr |_{\Lambda}$ defined by 
\begin{equation}
\label{deftau}
\tau(x,\sigma(x))(s_1(x)\otimes s_2(\sigma(x)))=
\frac{1}{h_{L_x}(e(x),e(x))}h_{L_x}(s_1(x),e(x))h_{L_{\sigma(x)}}(s_2(\sigma(x)),e(\sigma(x))), 
\end{equation}
for holomorphic sections $s_1,s_2$ of $L$, 
where $e$ is a holomorphic section of $L$, locally nonvanishing. It is clear that $\tau$ is a well defined global section 
of ${\mathcal{L}}^*\Bigr |_{\Lambda}\to\Lambda$ (it does not depend on the choice of $e$, because replacing $e(x)$ 
by $f(x)e(x)$, where $f(x)$ is a local non-vanishing holomorphic function, does not change the right hand side of (\ref{deftau}) 
due to (\ref{sigmareq})). Now let us explain why $\tau$ is covariant constant. 
We have already seen that with respect to a local holomorphic frame $e(x)$ in $L$ the connection form in $\iota^*({\mathcal{L}}\Bigr |_{\Lambda})$ 
is $d \log  h_{L_x}(e(x),e(x))$. Therefore the connection form in $\iota^*({\mathcal{L}} ^*\Bigr |_{\Lambda})$ 
is $-d \log  h_{L_x}(e(x),e(x))$. 
We get, using (\ref{sigmareq}) and (\ref{deftau}): 
$$
\nabla^* \tau (x,\sigma(x))(e(x)\otimes e(\sigma(x))=d\Bigl ( \tau (x,\sigma(x))(e(x)\otimes e(\sigma(x))) \Bigr ) 
$$
$$
-
\tau (x,\sigma(x))(e(x)\otimes e(\sigma(x)))d \log  h_{L_x}(e(x),e(x))
$$
$$
=d h_{L_x}(e(x),e(x))- h_{L_x}(e(x),e(x))d \log  h_{L_x}(e(x),e(x))=0.
$$
Let $d\mu=\pi_1^* dV$. 
For an $m$-dimensional submanifold $\Lambda'$ of $\Lambda$, with an $m$-form $d\mu'$ 
we have:   
$$
u_{\Lambda'}^{(k)}=\sum\limits_{j=1}^{d_k}\sum\limits_{l=1}^{d_k}
\varphi_j^{(k)}\otimes \varphi_l^{(k)} \int\limits_{\Lambda'}
\overline{\tau(x,\sigma(x))^{\otimes k} (\varphi_j^{(k)}(x)\otimes \varphi_l^{(k)}(\sigma(x)))}d\mu'(x,\sigma(x)).
$$ 
When $\Lambda'=\Lambda$ and $d\mu'=d\mu$, then    
$$
u_{\Lambda}^{(k)}=\sum\limits_{j=1}^{d_k}\sum\limits_{l=1}^{d_k}
\varphi_j^{(k)}\otimes \varphi_l^{(k)} \int\limits_{X}
\overline{\tau(x,\sigma(x))^{\otimes k} (\varphi_j^{(k)}(x)\otimes \varphi_l^{(k)}(\sigma(x)))}dV(x).
$$
\begin{remark}
Our way of associating semiclassical states to Bohr-Sommerfeld Lagrangian submanifolds 
is the same as the one, for example, in \cite{burns:10}. Instead, one could consider $\Lambda$ as 
$p(\tilde{\Lambda})$, where 
$p:P\to X\times X$ is the unit circle bundle in ${\mathcal{L}}^*\to X\times X $ 
(a principal $U(1)$-bundle, also 
a contact manifold), 
and  $\tilde{\Lambda}$ is a Legendrian submanifold of $P$ \cite{borthwick:95}, \cite{eliashberg:95}. 
Specifically, in Section \ref{sectantidiag} $\tilde{\Lambda}=\{ (x,\sigma(x),\tau(x,\sigma(x)))|\ x\in X\}$. 
\end{remark}
\begin{remark}
It would be interesting to interpret either the entropy of entanglement or the minimum distance 
to a separable vector as a number that characterizes how far $\Lambda$ is from being a product submanifold.  
\end{remark}
\begin{theorem} 
\label{entropyth} 
$$
\nu \Bigl ( \frac{u_{\Lambda}^{(k)}}{||u_{\Lambda}^{(k)}||}\Bigr ) = \ln d_k
$$
\end{theorem} 
\noindent {\bf Proof.} 
Let's write $u_{\Lambda}^{(k)}$ as a vector in the basis $\varphi_j^{(k)}\otimes \varphi_l^{(k)}$ and let's find 
$v_k=\frac{1}{||u_{\Lambda}^{(k)}||}u_{\Lambda}^{(k)}$ and $\rho_k=v_k\bar{v}_k^T$. We have: 
$$
u_{\Lambda}^{(k)}=\sum\limits_{j=1}^{d_k}\sum\limits_{l=1}^{d_k}
\varphi_j^{(k)}\otimes \varphi_l^{(k)} \int\limits_{X}
\overline{\tau(x,\sigma(x))^{\otimes k} (\varphi_j^{(k)}(x)\otimes \varphi_l^{(k)}(\sigma(x)))}dV(x), 
$$
$$
\tau(x,\sigma(x))^{\otimes k}(\varphi_j^{(k)}(x)\otimes \varphi_l^{(k)}(\sigma(x) ))
$$
$$
=\frac{1}{h_{L_x^{\otimes k}}(\varphi^{\otimes k}(x),\varphi^{\otimes k}(x))}h_{L_x^{\otimes k}}(\varphi_j^{(k)}(x),\varphi^{\otimes k}(x))
h_{L_{\sigma(x)}^{\otimes k}} ( \varphi_l^{(k)}(\sigma(x)),\varphi^{\otimes k}(\sigma(x))), 
$$
where $\varphi$ is a holomorphic section of $L$ (locally nonvanishing). Because of (\ref{sigmareq}) 
$$
\tau(x,\sigma(x))^{\otimes k}(\varphi_j^{(k)}(x)\otimes \varphi_l^{(k)}(\sigma(x) ))
$$
$$
=\frac{1}{h_{L_x^{\otimes k}}(\varphi^{\otimes k}(x),\varphi^{\otimes k}(x))}h_{L_x^{\otimes k}}(\varphi_j^{(k)}(x),\varphi^{\otimes k}(x))
\overline{h_{L_x^{\otimes k}}(\varphi_l^{(k)}(x),\varphi^{\otimes k}(x))}=
h_{L_x^{\otimes k}} 
(\varphi_j^{(k)}(x),\varphi_l^{(k)}(x)). 
$$
The last equality can be obtained by setting $\varphi_j^{(k)}(x)=a(x)\varphi^{\otimes k}(x)$, $\varphi_l^{(k)}(x)=b(x)\varphi^{\otimes k}(x)$. 
Hence
$$
u_{\Lambda}^{(k)}=\sum\limits_{j=1}^{d_k}\sum\limits_{l=1}^{d_k}
\varphi_j^{(k)}\otimes \varphi_l^{(k)} \langle \varphi_l^{(k)}, \varphi_j^{(k)}\rangle =
\sum\limits_{j=1}^{d_k}\varphi_j^{(k)}\otimes \varphi_j^{(k)}.
$$ 
Therefore in the basis $\varphi_1^{(k)}\otimes \varphi_1^{(k)}$, 
$\varphi_1^{(k)}\otimes \varphi_2^{(k)}$,...,$\varphi_1^{(k)}\otimes \varphi_{d_k}^{(k)}$,
$\varphi_2^{(k)}\otimes \varphi_1^{(k)}$, 
$\varphi_2^{(k)}\otimes \varphi_2^{(k)}$,...,$\varphi_2^{(k)}\otimes \varphi_{d_k}^{(k)}$,..., 
$\varphi_{d_k}^{(k)}\otimes \varphi_1^{(k)}$,$\varphi_{d_k}^{(k)}\otimes \varphi_2^{(k)}$,...,
$\varphi_{d_k}^{(k)}\otimes \varphi_{d_k}^{(k)}$ in $\Hk\otimes \Hk$ the vector 
$v_k$ is $\frac{1}{\sqrt{d_k}}\begin{pmatrix} e_1 \\ ... \\  e_{d_k}\end{pmatrix}$, where 
$\begin{pmatrix} e_1 &  ... &  e_{d_k}\end{pmatrix}=I_{d_k}$ is the $d_k\times d_k$ 
identity matrix. Then $\rho_k$ is 
$ \frac{1}{d_k} \sum\limits_{j=1}^{d_k}\sum\limits_{l=1}^{d_k}E_{jl}\otimes E_{jl}$, where 
$E_{jl}$ is the $d_k\times d_k$ matrix with the $jl$-th entry $1$ and the other entries equal to zero. 
From this we get: $\tr _2(\rho_k)= \frac{1}{d_k}\sum\limits_{j=1}^{d_k}E_{jj}=\frac{1}{d_k}I_{d_k}$.
It follows that $v_k$ is maximally entangled and $\nu (v_k)=\ln d_k$. 
$\Box$. 

Since the vectors $v_k=\frac{1}{||u_{\Lambda}^{(k)}||}u_{\Lambda}^{(k)}$ in Theorem \ref{entropyth} are maximally 
entangled, Corollary \ref{cormaxent} applies: 
\begin{corollary}
\label{cormindist}
The minimum distance from $v_k$ to a separable vector is  $\sqrt{\frac{d_k-1}{d_k}}=\sqrt{1-e^{-\nu(v_k)}}$. 
\end{corollary}

\section{Calculations on the sphere and on the torus}

\subsection{} 
\label{sphere}
Suppose $X=\Prj^n$ ($n\ge 1$), 
with the Fubini-Study K\"ahler form, $L$ is the hyperplane bundle, with the standard choice 
of a Hermitian metric. Denote by $\zeta_0,..., \zeta_n$ the homogeneous coordinates on $X$. 
The space $\Hk$ is usually identified with 
the space of homogeneous polynomials of degree $k$ in $\zeta_0,..., \zeta_n$. 
The antiholomorphic involution $X\to X$, 
$\sigma: \zeta=[\zeta_0:...:\zeta_n]\mapsto \sigma(\zeta)=[\bar{\zeta}_0:...:\bar{\zeta}_n]$, provides 
an antilinear map that sends the fiber $L^*_{\zeta}$ of the tautological bundle 
(i.e. the line $l$ through $0$ in $\C^{n+1}$ spanned by $\begin{pmatrix} \zeta_0 \\ ...\\ \zeta_n\end{pmatrix} $)   
to $\bar{l}=\{ w\in \C^{n+1}| \ \bar{w}\in l\}$. 
Theorem \ref{entropyth} says that for the Lagrangian states associated to $\Prj^n$ antidiagonally embedded 
into  $\Prj^n\times \Prj^n$ the entropy of entanglement is $\ln {\mbox{dim}} \Hk =\ln \binom{n+k}{n}$. 

Everything can be written very explicitly for $n=1$. On $\Prj ^1$ we use the homogeneous coordinates $[\zeta_0:\zeta_1]$. 
The space $\Hk$ of holomorphic sections of $L^{\otimes k}$, $k\in\N$,  is  the space of homogeneous degree $k$ polynomials 
in $\zeta_0$, $\zeta_1$. 
On the affine chart $\U= \{ [\zeta_0:\zeta_1]\in \Prj ^1 | \zeta_1\ne 0\}\cong \C$ homogeneous degree $k$ polynomials 
in $\zeta_0$, $\zeta_1$ become polynomials in $z=\zeta_0/\zeta_1$ of degree $\le k$, and  
$\Hk$
is identified with the space of polynomials in complex variable $z$ of degree $\le k$, with the inner product 
$$
\langle \phi,\psi\rangle = \frac{i}{2\pi }\int\limits_{\C }\frac{\phi(z)\overline{\psi(z)}}{(1+|z|^2)^{k+2}}dzd\bar{z} 
$$
(\cite{marche:15} 4.1.1. or \cite{bloch:03} 3.1). $\Prj^1\cong \U \cup \{ [1:0]\}$. 
An orthonormal basis in $\Hk$ is 
$$
\varphi_j^{(k)}=\sqrt{\frac{(k+1)!}{j!(k-j)!}}z^j, \ j=0,...,k 
$$
(this is easily verified by a calculation in polar coordinates on $\R^2$). Note that  
$\overline{\varphi_j^{(k)}(\sigma(z))}=\overline{\varphi_j^{(k)}(\bar{z})}=\varphi_j^{(k)}(z)$. 
Denote ${\mathbf{z}}=[z:1]\in \U$.  
The Hermitian metric on $L^*$ is given by 
$$
h_{L_{\mathbf{z}}^*}\Bigl (\begin{pmatrix} z \\  1\end{pmatrix}, \begin{pmatrix} z \\  1\end{pmatrix} \Bigr ) =1+|z|^2,
$$
which implies that for $f,g\in\Hk$ 
$$
h_{L_{\mathbf{z}}^{\otimes k}}(f(z),g(z))=\frac{f(z)\overline{g(z)}}{(1+|z|^2)^k}.
$$
In particular, 
$$
h_{L_{\mathbf{z}}^{\otimes k}}(\varphi_j^{(k)},\varphi_l^{(k)})=
\frac{(k+1)!}{\sqrt{j!(k-j)!l!(k-l)!}}\frac{z^j\bar{z}^l}{(1+|z|^2)^k}.
$$
The section $\tau$ is $\tau(z,\bar{z})=\frac{1}{1+|z|^2} \begin{pmatrix} z \\  1\end{pmatrix} \otimes 
\begin{pmatrix} \bar{z} \\  1\end{pmatrix}$. We have: 
$$
\tau(z,\bar{z})^{\otimes k} (\varphi_j^{(k)}(z)\otimes \varphi_l^{(k)}(\sigma(z))=
\frac{1}{(1+|z|^2)^k}\varphi_j^{(k)}(z) \varphi_l^{(k)}(\bar{z})=
h_{L_{\mathbf{z}}^{\otimes k}}(\varphi_j^{(k)}(z),\varphi_l^{(k)}(z)). 
$$
$\tau$ is covariant constant, as it is shown in Section \ref{sectantidiag}.  
$$
u_{\tau ({\mathbf{z}},{\mathbf{\bar{z}}})}^{(k)}( {\mathbf{w}},{\mathbf{\bar{w}}})
=\sum\limits_{j=0}^{k}\sum\limits_{l=0}^{k} \frac{
\overline{\varphi_j^{(k)}(z) \varphi_l^{(k)}(\bar{z})} }
{(1+|z|^2)^k} 
\varphi_j^{(k)}(w)\otimes \varphi_l^{(k)}(\bar{w}), 
$$
and 
$$
u_{\Lambda}^{(k)}=\sum\limits_{j=0}^{k}\sum\limits_{l=0}^{k}
\varphi_j^{(k)}\otimes \varphi_l^{(k)} \int\limits_{\C}
\frac{
\overline{\varphi_j^{(k)}(z) \varphi_l^{(k)}(\bar{z})} }
{(1+|z|^2)^k} \frac{i}{2\pi}\frac{dz d\bar{z}}{(1+|z|^2)^2} 
$$
$$
=\sum\limits_{j=0}^{k}\sum\limits_{l=0}^{k}
\varphi_j^{(k)}\otimes \varphi_l^{(k)} \int\limits_{\C}
\frac{\varphi_l^{(k)}(z)\overline{\varphi_j^{(k)}(z)} }{(1+|z|^2)^k }  
\frac{i}{2\pi}\frac{dz d\bar{z}}{(1+|z|^2)^2}
$$ 
$$
=\sum\limits_{j=0}^{k}\sum\limits_{l=0}^{k}
\varphi_j^{(k)}\otimes \varphi_l^{(k)} 
\langle \varphi_l^{(k)}, \varphi_j^{(k)}\rangle=
\sum\limits_{j=0}^{k}
\varphi_j^{(k)}\otimes \varphi_j^{(k)}.
$$
Therefore $u_{\Lambda}^{(k)}$ in the basis $\varphi_0^{(k)}\otimes \varphi_0^{(k)}$, 
$\varphi_0^{(k)}\otimes \varphi_1^{(k)}$,...,$\varphi_0^{(k)}\otimes \varphi_k^{(k)}$,
$\varphi_1^{(k)}\otimes \varphi_0^{(k)}$, 
$\varphi_1^{(k)}\otimes \varphi_1^{(k)}$,...,$\varphi_1^{(k)}\otimes \varphi_k^{(k)}$,..., 
$\varphi_k^{(k)}\otimes \varphi_0^{(k)}$,$\varphi_k^{(k)}\otimes \varphi_1^{(k)}$,...,$\varphi_k^{(k)}\otimes \varphi_k^{(k)}$ in $\Hk\otimes \Hk$ 
is $\begin{pmatrix} e_1 \\ e_2 \\... \\  e_{k+1}\end{pmatrix}$, where 
$\begin{pmatrix} e_1 & e_2 & ... &  e_{k+1}\end{pmatrix}=I_{k+1}$ is the $(k+1)\times(k+1)$ 
identity matrix. We get: $v_k=\frac{1}{||u_{\Lambda}^{(k)}||}u_{\Lambda}^{(k)}=
\frac{1}{\sqrt{k+1}}\begin{pmatrix} e_1 \\ e_2 \\... \\  e_{k+1}\end{pmatrix}$, 
$v_k\bar{v}_k^T$ is 
$ \frac{1}{k+1} \sum\limits_{j=1}^{k+1}\sum\limits_{l=1}^{k+1}E_{jl}\otimes E_{jl}$, 
$\tr _2(v_k\bar{v}_k^T)= \frac{1}{k+1}\sum\limits_{j=1}^{k+1}E_{jj}=\frac{1}{k+1}I_{k+1}$.
It follows that $v_k$ is maximally entangled and $\nu (v_k)=\ln (k+1)$. 

The Schmidt decomposition of $v_k$ is $v_k=\sum\limits_{j=1}^{k+1} \frac{1}{\sqrt{k+1}}e_j\otimes e_j$, 
and $u_s^{(k)}=\frac{1}{\sqrt{k+1}}e_1\otimes e_1$. 
The distance 
to a closest separable vector $D(v_k,u_s^{(k)})=\sqrt{\frac{k}{k+1}}=\sqrt{1-e^{-\nu(v_k)}}$.

\begin{remark}
In general, a closest separable vector is not unique. 
Here, for example,  
$D(v_k, \frac{1}{\sqrt{k+1}}e_2\otimes e_2)=\sqrt{\frac{k}{k+1}}$ too. 
\end{remark}

Let us now calculate the entropy of entanglement 
for the states associated to $\Lambda'= 
\{ ({\mathbf{z}},{\mathbf{\bar{z}}})\in\Lambda \ | \  z=e^{i\Theta}, \ 0\le \Theta < 2\pi\}$ with $d\mu=d\Theta$. 
$$
u_{\Lambda '}^{(k)}=\sum\limits_{j=0}^{k}\sum\limits_{l=0}^{k}
\varphi_j^{(k)}\otimes \varphi_l^{(k)} \int\limits_0^{2\pi}
\frac{
\overline{\varphi_j^{(k)}(z) \varphi_l^{(k)}(\bar{z})} }
{(1+|z|^2)^k} \Bigr |_{z=e^{i\Theta}} d\Theta=
\sum\limits_{j=0}^{k} \frac{\pi}{2^{k-1}}\frac{(k+1)!}{j!(k-j)!}
\varphi_j^{(k)}\otimes \varphi_j^{(k)}. 
$$
Using the identity $\sum\limits_{j=0}^k\binom{k}{j} ^2=\binom{2k}{k}$, we get: 
$$
v_k=\frac{1}{||u_{\Lambda'}^{(k)}||}u_{\Lambda'}^{(k)}=
\frac{k!}{\sqrt{(2k)!}}\begin{pmatrix} \binom{k}{0}e_1 \\ \binom{k}{1}e_2 \\... \\  \binom{k}{k}e_{k+1}\end{pmatrix},
$$ 
$$
v_k\bar{v}_k^T=
\frac{(k!)^2}{(2k)!} \sum\limits_{j=1}^{k+1}\sum\limits_{l=1}^{k+1}\binom{k}{j-1}\binom{k}{l-1}E_{jl}
\otimes E_{jl}, 
$$
$$
\tr _2(v_k\bar{v}_k^T)= 
\frac{(k!)^2}{(2k)!} \sum\limits_{j=1}^{k+1}\binom{k}{j-1}^2E_{jj}=
\frac{(k!)^2}{(2k)!}\begin{pmatrix} \binom{k}{0} ^2 & & & \\ 
& \binom{k}{1} ^2 & & \\
 & &  ... & \\
 & & & \binom{k}{k} ^2\end{pmatrix} ,
$$  
$$
\nu(v_k)=-\frac{(k!)^2}{(2k)!} \sum\limits_{j=0}^{k}\binom{k}{j} ^2\ln 
\Bigl [ \frac{(k!)^2}{(2k)!}\binom{k}{j} ^2\Bigr ] .
$$
We conclude that the vectors $v_k$ are not maximally entangled. Also, they are clearly not separable. 

\subsection{} 
\label{torus}
Another source of examples of complex manifolds with real structures is  
compact Riemann surfaces. In genus $1$ those are   
elliptic curves with a real $j$-invariant (biholomorphic to 
$\C/\Gamma$ where $\Gamma=\Z+w\Z$ with $w$ either on the boundary of the region 
$\{ z\in \C| \ |z|\ge 1, |Re(z)|\le 1/2\}$ or $w=it$ with $t\ge 1$) \cite{alling:81}. 
For genus $g\ge 2$ there is extensive literature, - see, for example, an account given in \cite{bujalance:01}.

Let $X$ be the torus $\C/\Gamma$ where $\Gamma=\Z+i\Z$. 
Let $\mu\in \R$, $k\in \N$, we will assume $k\ge 3$. Recall that a $k$-th order 
theta-function with characteristics $(\mu,0)$ for this lattice 
is an entire function $f:\C\to \C$ such that 
$$
f(z+m+in)=e^{2\pi i(-\frac{k}{2}(in^2+2nz)+m\mu)}  f(z) 
$$
for all $m,n\in\Z$, $z\in\C$, the space $\Theta (k;\mu,0)$ of these theta functions is $k$-dimensional, and the functions 
$$
\theta_j^{(k,\mu)}(z)=\sum\limits_{n\in\Z}e^{2\pi i [\frac{1}{2}ki (n+\frac{\mu+j}{k})^2+(n+\frac{\mu+j}{k})kz ]}
 \ , \ j=1,...,k
$$
form  a basis in $\Theta (k;\mu,0)$ \cite{baily:65}. 

Let $L_{k,\mu}\to X$ 
be the hermitian holomorphic line bundle on $X$ such that $H^0(X,L_{k,\mu})\cong \Theta (k;\mu,0)$ 
under the usual identification between holomorphic sections of a line bundle on a torus 
and theta-functions (see e.g. \cite{debarre:05}), with the Hermitian metric determined by the function
$h(z)=e^{\frac{k\pi}{4}(z-\bar{z})^2}$, so that the inner product on $H^0(X,L_{k,\mu})\cong \Theta (k;\mu,0)$ is 
$$
\langle f,g\rangle = \int\limits_0^1\int\limits_0^1 f(z)\overline{g(z)}
e^{\frac{k\pi}{2}(z-\bar{z})^2}\ dx \ dy. 
$$
A straightforward calculation shows that the basis $\{ \theta_j^{(k,\mu)}(z)\}$
is orthogonal. 
Denote $\varphi_j^{(k,\mu)}=\frac{1}{||\theta_j^{(k,\mu)}||}\theta_j^{(k, \mu)}$, $j=1,...,k$. 
Take $\sigma:z\mapsto -\bar{z}$ as the antiholomorphic involution. Note that  
$\overline{\varphi_j^{(k)}(\sigma(z))}=\varphi_j^{(k)}(z)$.  

Define a section $\tau_k$ of 
$\iota^*((\pi_1^*L_{k,\mu} \otimes \pi_2^*L_{k,\mu})^*\Bigr | _{\Lambda})\cong L_{k,\mu}^*\otimes L_{k,\mu}$ 
by  
$$
\tau_k(z,-\bar{z})(s_1(z)\otimes s_2(-\bar{z}))=
e^{\frac{k\pi}{2}(z-\bar{z})^2}s_1(z) s_2(-\bar{z}) 
$$
(note that the expression in the right hand side is $\Gamma$-invariant). 
\begin{remark}
In this section we slightly change notations. Previously we defined a covariantly constant section $\tau$ of  ${\mathcal{L}}^*\Bigr | _{\Lambda}$
and used $\tau^{\otimes k}$ for dealing with $L^{\otimes k}$. In this section, because of theta characteristics, 
the role of $L^{\otimes k}$ is played by $L_1^{\otimes k}\otimes L_2$ for certain line bundles $L_1$, $L_2$. For this reason 
we introduce and use $\tau_k$ instead of $\tau$ and $\tau^{\otimes k}$.    
\end{remark}
We have: 
$$
\tau_k(z,-\bar{z}) (\varphi_j^{(k,\mu)}(z)\otimes \varphi_l^{(k,\mu)}(\sigma(z)))=
e^{\frac{k\pi}{2}(z-\bar{z})^2}\varphi_j^{(k,\mu)}(z) \varphi_l^{(k,\mu)}(-\bar{z})=
e^{\frac{k\pi}{2}(z-\bar{z})^2}\varphi_j^{(k,\mu)}(z)\overline { \varphi_l^{(k,\mu)}(z)}.
$$
$\tau_k$ is covariant constant (see Section \ref{sectantidiag}). 
$$
u_{\Lambda}^{(k)}=\sum\limits_{j=1}^{k}\sum\limits_{l=1}^{k}
\varphi_j^{(k,\mu)}\otimes \varphi_l^{(k,\mu)} \int\limits_0^1\int\limits_0^1
\overline{\tau_k(z,-\bar{z}) (\varphi_j^{(k,\mu )}(z)\otimes \varphi_l^{(k,\mu )}(-\bar{z}))}dxdy
$$
$$
=\sum\limits_{j=1}^{k}\sum\limits_{l=1}^{k}
\varphi_j^{(k,\mu)}\otimes \varphi_l^{(k,\mu)} \int\limits_0^1\int\limits_0^1
\overline{\varphi_j^{(k,\mu)}(z)}\varphi_l^{(k,\mu)}(z)e^{ k\frac{\pi}{2}(z-\bar{z})^2 }dxdy=
\sum\limits_{j=1}^{k}\varphi_j^{(k,\mu)}\otimes \varphi_j^{(k,\mu)},  
$$
therefore $u_{\Lambda}^{(k)}$ in the basis  $\varphi_1^{(k,\mu)}\otimes \varphi_1^{(k,\mu)}$,  
$\varphi_1^{(k,\mu)}\otimes \varphi_2^{(k,\mu)}$, ..., $\varphi_1^{(k,\mu)}\otimes \varphi_k^{(k,\mu)}$, ... ,
$\varphi_k^{(k,\mu)}\otimes \varphi_1^{(k,\mu)}$, ... ,$\varphi_k^{(k,\mu)}\otimes \varphi_k^{(k,\mu)}$,  
in $\Theta (k;\mu,0)\otimes \Theta (k;\mu,0)$ is $\begin{pmatrix} e_1 \\ e_2 \\... \\  e_{k}\end{pmatrix}$, 
$$
v_k=\frac{1}{||u_{\Lambda}^{(k)}||}u_{\Lambda}^{(k)}=
\frac{1}{\sqrt{k}}\begin{pmatrix} e_1 \\ e_2 \\... \\  e_{k}\end{pmatrix}
$$
and by a procedure similar to the one used in the proof of Theorem \ref{entropyth} or in Section \ref{sphere}, 
we get: 
$\tr _2(v_k\bar{v}_k^T)= \frac{1}{k}I_{k}$, vectors $v_k$ are maximally entangled, $\nu (v_k)=\ln k$. 

The Schmidt decomposition of $v_k$ is $v_k=\sum\limits_{j=1}^{k} \frac{1}{\sqrt{k}}e_j\otimes e_j$, 
and $u_s^{(k)}=\frac{1}{\sqrt{k}}e_1\otimes e_1$.
The distance from $v_k$ 
to a closest separable vector $D(v_k,u_s^{(k)})=\sqrt{\frac{k-1}{k}}=\sqrt{1-e^{-\nu(v_k)}}$. 

{\bf Acknowledgments.} We are thankful to N. Johnston, Y. Karshon, J. Rosenberg, A. Uribe, for discussions, and to the referees and the editors 
for useful suggestions that helped improve the exposition in the paper.

\end{document}